\documentclass[sigconf]{acmart}
\usepackage{booktabs} % For formal tables

\usepackage{multirow}
\usepackage{amsmath}
\usepackage{subfig}
\usepackage{enumitem}

\usepackage{multicol}
\usepackage{multirow}
\usepackage{float}

\usepackage{hhline}
\usepackage{booktabs} % For formal tables
\usepackage{amssymb}
\usepackage{graphicx}
\usepackage{textcomp}
\usepackage[labelfont=bf]{caption}

\usepackage{algorithm}
\usepackage{algpseudocode}

\def\BibTeX{{\rm B\kern-.05em{\sc i\kern-.025em b}\kern-.08emT\kern-.1667em\lower.7ex\hbox{E}\kern-.125emX}}
    
\copyrightyear{2019}
\acmYear{2019} 
\acmConference[WWW '19 Companion]{Companion Proceedings of the 2019 World Wide Web Conference}{May 13--17, 2019}{San Francisco, CA, USA}
\begin{document}

%
% The "title" command has an optional parameter, allowing the author to define a "short title" to be used in page headers.
\title{Less is More: Semi-Supervised Causal Inference for Detecting Pathogenic Users in Social Media}

%
% The "author" command and its associated commands are used to define the authors and their affiliations.
% Of note is the shared affiliation of the first two authors, and the "authornote" and "authornotemark" commands
% used to denote shared contribution to the research.

\author{Hamidreza Alvari, Elham Shaabani, Soumajyoti Sarkar, Ghazaleh Beigi, Paulo Shakarian }
\affiliation{\institution{Arizona State University}}
\email{{halvari,eshaaban,ssarka18,gbeigi,shak}@asu.edu}

%
% By default, the full list of authors will be used in the page headers. Often, this list is too long, and will overlap
% other information printed in the page headers. This command allows the author to define a more concise list
% of authors' names for this purpose.
\renewcommand{\shortauthors}{Alvari et al.}
\renewcommand{\shorttitle}{Semi-Supervised Causal Inference for Detecting Pathogenic Users}

%
% The abstract is a short summary of the work to be presented in the article.
\begin{abstract}
Recent years have witnessed a surge of manipulation of public opinion and political events by malicious social media actors. These users are referred to as ``Pathogenic Social Media (PSM)" accounts. PSMs are key users in spreading misinformation in social media to viral proportions. These accounts can be either controlled by real users or automated bots. Identification of PSMs is thus of utmost importance for social media authorities. The burden usually falls to automatic approaches that can identify these accounts and protect social media reputation. However, lack of sufficient labeled examples for devising and training sophisticated approaches to combat these accounts is still one of the foremost challenges facing social media firms. In contrast, unlabeled data is abundant and cheap to obtain thanks to massive user-generated data. In this paper, we propose a semi-supervised causal inference PSM detection framework, \textsc{SemiPsm}, to compensate for the lack of labeled data. In particular, the proposed method leverages unlabeled data in the form of manifold regularization and only relies on cascade information. This is in contrast to the existing approaches that use exhaustive feature engineering (e.g., profile information, network structure, etc.). Evidence from empirical experiments on a real-world ISIS-related dataset from Twitter suggests promising results of utilizing unlabeled instances for detecting PSMs.     
\end{abstract}

%
% The code below is generated by the tool at http://dl.acm.org/ccs.cfm.
% Please copy and paste the code instead of the example below.
%

% \ccsdesc[500]{Computer systems organization~Embedded systems}
% \ccsdesc[300]{Computer systems organization~Redundancy}
% \ccsdesc{Computer systems organization~Robotics}
% \ccsdesc[100]{Networks~Network reliability}

%
% Keywords. The author(s) should pick words that accurately describe the work being
% presented. Separate the keywords with commas.

\maketitle

\section{Introduction}
Over the past years, social media play major role in massive dissemination of misinformation online. Political events and public opinion on the Web and social networks have been allegedly manipulated by different forms of accounts including real users and automated software (a.k.a social bots or sybil accounts). Pathogenic Social Media (PSM) accounts are among those that are responsible for such a massive spread of disinformation online and swaying normal users' opinion~\cite{alvari2018early,shaabani2019psm}. These accounts (1) are usually owned by either normal users or automated bots, (2) seek to promote or degrade certain ideas; and (3) can appear in many forms such as terrorist supporters (e.g., ISIS supporters), water armies or fake news writers. Understanding the behavior of PSMs will allow social media to take countermeasures against their propaganda at the early stage and reduce their threat to the public. 

The problem of identification of PSMs has long been addressed in the past by the research community mostly in the form of bot detection. Several approaches especially supervised learning methods have been proposed in the literature and they have shown promising results~\cite{kudugunta2018deep}. However, for the most part, these approaches are often based on labeled data and exhaustive feature engineering. Examples of such feature groups include but are not limited to content, sentiment of posts, profile information and network features. These approaches are thus very expensive as they require significant amount of efforts to design features and annotate large labeled datasets. In contrast, unlabeled data is ubiquitous and cheap to collect thanks to the massive user-generated data produced on a daily basis. Thus, in this work we set out to examine if unlabeled instances can be utilized to compensate for the lack of enough labeled data.

\textbf{Present Work.} In this paper, semi-supervised causal inference is tailored to detect PSMs who are promoters of misinformation online. We cast the problem of identifying PSMs as an optimization problem and propose a semi-supervised causal learning framework which utilizes unlabeled examples through manifold regularization~\cite{belkin2006manifold}. In particular, we incorporate causality-based features extracted from users' activity log (i.e., cascades of retweets) as regularization terms into the optimization problem. In this work, causal inference is leveraged in an effort to capture whether or not PSMs exert causal influences while making a message viral. %: (1) causality-based features; and (2) LSTM-based features. 
Our causality-based features are built upon \textit{Suppes' theory of probabilistic causation}~\cite{suppes1970} whose central concept is \textit{prima facie causes}: an event to be recognized as a cause, must occur before the effect and must lead to an increase of the likelihood of observing the effect. While there exists a prolific literature on causality and their great impact in the computer-science community (see~\cite{Pearl:2009:CMR:1642718} for instance), we build our foundation on \textit{Suppes' theory} as it is computationally less complex.  %The LSTM-based features are extracted via Long Short Term Memory (LSTM)~\cite{hochreiter1997long}, a superior variant of Recurrent Neural Networks (RNNs) which have shown promising results for different tasks~\cite{.}.

\textbf{Key idea and highlights.} To summarize, this paper makes the following main contributions:
\begin{itemize}
	\item We frame the problem of detecting PSM accounts as an optimization problem and present a Laplacian semi-supervised causal inference \textsc{SemiPsm} for solving it. The unlabeled data are utilized via manifold regularization. 
	%\item We examine causality-based and LSTM-based features extracted from cascades of retweets of a real-world ISIS-related dataset from Twitter.
	\item Manifold regularization used in the resultant optimization formulation is built upon causality-based features created on a notion of \textit{Suppes' theory of probabilistic causation}.
	\item We conduct a suite of experiments using different supervised and semi-supervised methods. Empirical experiments on a real-world ISIS-related dataset from Twitter suggests the effectiveness of the proposed semi-supervised causal inference over the existing methods. 
\end{itemize}

The remainder of the paper is organized as follows. In Section 2, we present the proposed framework. Section 3 summarizes the empirical experiments on an ISIS-related dataset from Twitter. In Section 4 we review the state-of-the-art methods. We conclude the paper in Section 5 by presenting the future work.

\section{The Proposed Method}
In this section, we first provide the causal inference used to extract features out of users' activity log. Then, we detail the proposed semi-supervised causal inference, namely \textsc{SemiPsm}, for detecting PSM accounts. 
\subsection{Causal Inference}
%\begin{table*}[t]\small
%	\centering
%	\caption{Frequently used symbols.}
%\begin{tabular}{|p{.8cm}|p{6cm}|}
%	\begin{tabular}{|c|l|}
%		\cline{1-2}
%		\textbf{Name}  & \textbf{Definition} \\
%		\hhline{==}
%		$\mathbf{A}$ & Action log (i.e., activity log) of the form \textit{Actions(User,Action,Time)}. \\ \cline{1-2}
%		$\mathbf{C}$ & Communities over graph $\mathbf{G}$.  \\ \cline{1-2}
%		$\mathbf{A}_m$ & Cascade related to a given message $m$  \\ \cline{1-2}
%		$\mathbf{M}_{vir}$ & Set of all viral messages  \\ \cline{1-2}
%		$\rho$ & Probability of a message going viral.  \\ \cline{1-2}
%		$\rho_u$ & Probability of a message going viral given a key user $u$ has participated in.  \\ \cline{1-2}
%		$p_{i,j}$ & Probability that key users $i$ and $j$ make a message viral and $i$ precedes $j$.  \\ \cline{1-2}
%		$p_{\neg i,j}$ & Probability that only key user $j$ make a message viral \\ \cline{1-2}
%		$\mathcal{P}_{i,j}$ & $p_{i,j}$ over $[t_0,t]$  \\ \cline{1-2}
%		$\mathcal{P}_{\neg i,j}$ & $p_{\neg i,j}$ over $[t_0,t]$  \\ \cline{1-2}
%		$\mathcal{R}(i)$ &   Set of related users for user $i$ over $[t_0,t]$.\\ \cline{1-2}
%		$\mathcal{Q}(j)$ &   Set of all users over $[t_0,t]$ that user $j$ belongs to their related users.\\ \cline{1-2}
%	\end{tabular}
%	\label{tb:symbols}
%\end{table*}
%In this paper, we use bold uppercase letters (e.g., $\mathbf{A}$) to denote sets, bold lowercase letters (e.g., $\mathbf{x}$) to denote vectors,  calligraphic uppercase letters (e.g., $\mathcal{P}$) to denote quantities over a time period, and lowercase letters (e.g., $\rho$) to denote scalars. 

We follow the convention of~\cite{Goyal:2010} and assume an \textit{action log} $\mathcal{A}$ of the form \textit{Actions(User,Action,Time)}, which contains tuples $(i,a_i,t_i)$ indicating that user $i$ has performed action $a_i$ at time $t_i$. For ease of exposition, we slightly abuse the notation and use the tuple $(i,m,t)$ to indicate that user $i$ has posted (tweeted/retweeted) message $m$ at time $t$. For a given message $m$ we define a \textit{cascade} of actions as $\mathcal{A}_m= \{(i,m',t)\in\mathcal{A}|m'=m\}$. User $i$ is called $m$-participant if there exists $t_i$ such that $(i,m,t_i)\in\mathcal{A}$. Users who have adopted a message in the early stage of its life span are called \textit{key users}:

%\begin{description}
\textbf{Definition 1 (Key Users).} \textit{Given message $m$, $m$-participant $i$ and cascade $\mathcal{A}_m$, we say user $i$ is a key user iff user $i$ precedes at least $\phi$ fraction of other $m$-participants where $\phi\in(0,1)$. In other words,} $|\mathcal{A}_m|\times\phi \leq |\{j|\exists t': (j,m,t')\in \mathcal{A}\wedge t < t'\}|$, \textit{where $|.|$ is the cardinality of a set}. 
%\end{description}\label{def:key_users}

Next, we shall define viral messages as follows.

%\begin{description}
\textbf{Definition 2 (Viral Messages).} \textit{Given a threshold $\theta$, we say a message $m\in \mathbf{M}$ is viral iff $|\mathcal{A}_m| \geq \theta$. We denote a set of all viral messages by $\mathbf{M}_{vir}$.}
%\end{description}

The prior probability of a message going viral is $\rho = |\mathbf{M}_{vir}|/|\mathbf{M}|$. The probability of a message going viral given key user $i$ has participated in, is computed as follows:

\begin{equation}
\rho_i = \frac{|\{m|m\in \mathbf{M}_{vir}\wedge \textit{i is a key user}\}|}{|\{m|m\in \mathbf{M} \wedge \textit{i is a key user}\}|}
\end{equation}

The probability that key users $i$ and $j$ tweet/retweet message $m$ chronologically and make it viral, is computed as:

\begin{equation}
p_{i,j}=\frac{|\{m\in \mathbf{M}_{vir}|\exists t, t': t<t' \wedge (i,m,t), (j,m,t') \in \mathcal{A}\}|}{|\{m\in \mathbf{M}|\exists t, t': t<t' \wedge (i,m,t), (j,m,t') \in \mathcal{A}\}|}
\end{equation}

To examine how causal user $i$ was in helping a message $m$ going viral, we shall explore what will happen if we exclude user $i$ from $m$. We define the probability that \textit{only} key user $j$ has made a message $m$ viral, i.e. user $i$ has not posted $m$ or does not precede $j$ as: 

\begin{equation}
p_{\neg i,j}=
\frac{|\{m\in \mathbf{M}_{vir}|\exists t': (j,m,t') \in \mathcal{A} \wedge \nexists t:t< t', (i,m,t) \in \mathcal{A}\}|}{|\{m\in \mathbf{M}|\exists t': (j,m,t') \in \mathcal{A} \wedge \nexists t:t< t', (i,m,t) \in \mathcal{A}\}|}
\end{equation}

In this work we adopt the notion of \textit{prima facie causes} which is at the core of Suppes' theory of probabilistic causation~\cite{suppes1970} and introduce causality metrics. According to this theory, \textit{a certain event to be recognized as a cause, must occur before the effect and must lead to an increase of the likelihood of observing the effect}. Accordingly, prima facie causal users are defined as follows:

%\begin{description}
\textbf{Definition 3 (Prima Facie Causal Users).} \textit{A user $i$ is said to be Prima Facie causal user for cascade $\mathcal{A}_m$ iff: (1) user $i$ is a key user of $m$, (2) $m \in \mathbf{M}_{vir}$, and (3) $\rho_i>\rho$.
}
%\end{description}

%To seek if a given Prima Facie causal user is causal, 
We use the concept of \textit{related users} from a rule-based system~\cite{DBLP:journals/corr/StantonTJVCS15} which was an extension to the causal inference framework in~\cite{DBLP:journals/corr/abs-1205-2634}. Accordingly, we call users $i$ and $j$ $m$-related if (1) they are Prima Facie causal users for $m$, and (2) $i$ precedes $j$. We then define a set of user $i$'s related users as $\mathbf{R}(i) = \{j|j\ne i \textit{ and } i, j \textit{ are \textit{m}-related}\}$.

In this work, we use the time-decay causal metrics introduced in~\cite{alvari2018early} which are built on Suppes' theory. %These metrics are fed to the final semi-supervised causal inference framework, as features. 
The first metric used in this work is $\mathcal{E}_{K\&M}$ which is computed over a given time interval $I$ as follows: 

\begin{equation}
\mathcal{E}^I_{K\&M}(i)=\frac{ \sum_{j\in \mathcal{R}(i)}(\mathcal{P}_{i,j}-\mathcal{P}_{\neg i,j}) }{|\mathcal{R}(i)|}
\end{equation}

\noindent where $\mathcal{R}(i)$, $\mathcal{P}_{i,j}$, and $\mathcal{P}_{\neg i,j}$ are now defined over $I$. This metric estimates the causality score of user $i$ in making a message \textit{viral}, by taking the average of $\mathcal{P}_{i,j}-\mathcal{P}_{\neg i,j}$ over $\mathbf{R}(i)$. The intuition here is that user $i$ is more likely to be a cause of message $m$ to become viral than user $j$, if $\mathcal{P}_{i,j}-\mathcal{P}_{\neg i,j} > 0$. This metric cannot spot all PSMs, hence another metric is defined, namely relative likelihood causality $\mathcal{E}_{rel}$. This metric works by assessing relative difference between $\mathcal{P}_{i,j}$, and $\mathcal{P}_{\neg i,j}$:

\begin{equation}
\mathcal{E}^I_{rel}(i)=\frac{ \mathcal{S}(i,j) }{|\mathcal{R}(i)|}
\end{equation}

\noindent where $\mathcal{S}(i,j)$ is defined as follows and $\alpha$ is infinitesimal:

\begin{equation}
\mathcal{S}(i,j)=\begin{cases}
\frac{\mathcal{P}_{i,j}}{\mathcal{P}_{\neg i,j} + \alpha}-1, & \mathcal{P}_{i,j} > \mathcal{P}_{\neg i,j}\\
%0, & \mathcal{P}_{i,j} = \mathcal{P}_{\neg i,j} \\
1- \frac{\mathcal{P}_{\neg i,j}}{\mathcal{P}_{i,j}}, & \mathcal{P}_{i,j} \leq \mathcal{P}_{\neg i,j}
\end{cases}
\end{equation}

Two other neighborhood-based metrics were also defined in~\cite{alvari2018early}, first of which is computed as:

\begin{equation}
\mathcal{E}^I_{nb}(j)=\frac{ \sum_{i\in \mathcal{Q}(j)}\mathcal{E}^I_{K\&M}(i) }{|\mathcal{Q}(j)|}
\end{equation}

\noindent where $\mathcal{Q}(j)=\{i|j\in \mathcal{R}(i)\}$ is the set of all users that user $j$ belongs to their related users sets.
%\begin{equation}\small
%	\mathcal{Q}(j)=\{i|j\in \mathcal{R}(i)\}
%\end{equation} 
Similarly, the second metric is the weighted version of the above metric and is called weighted neighborhood-based causality and is calculated as:

\begin{equation}
\mathcal{E}^I_{wnb}(j)=\frac{ \sum_{i\in \mathcal{Q}(j)}w_i\times\mathcal{E}^I_{K\&M}(i) }{\sum_{i\in \mathcal{Q}(j)}w_i}
\end{equation}

The aim of this metric is to capture different impacts that users in $Q(j)$ might have on user $j$. 

%Previous causal metrics do not take into account time-decay effect. They assume a steady trend for computing causality scores. This is an unrealistic assumption, as causality of users may change over time. We introduce a generic decay-based metric. Our metric assigns different weights to different time points of a given time interval, inversely proportional to their distance from $t$ (i.e., smaller distance is associated with higher weight). Specifically, it performs the following: it (1) breaks down the given time interval into shorter time periods using a sliding time window, (2) deploys an exponential decay function of the form $f(x)=e^{-\alpha x}$ to account for the time-decay effect, and (3) takes average of the causality values computed over each sliding time window. Formally, $\xi^I_{k}$ is defined as follows, where $k\in\{K\&M,rel,nb,wnb\}$:   

\subsection{Final set of Features} Finally, the causal metrics discussed in the previous section will be fed as features to the semi-supervised framework-- this will be described in the next section. The final set of features is in the following generic form $\xi^I_{k}$ where $k\in\{K\&M,rel,nb,wnb\}$~\cite{alvari2018early}: 

\begin{equation}\small
\xi_{k}^I(i)=\frac{1}{|\mathcal{T}|}\sum_{t'\in \mathcal{T}}e^{-\sigma (t-t')}\times\mathcal{E}^{\Delta}_{k}(i)
\end{equation}

Here, $\sigma$ is a scaling parameter of the exponential decay function, $\mathcal{T}=\{t'|t'=t_0+j\times \delta, j\in\mathbb{N} \wedge t'\leq t-\delta\}$ is a sequence of sliding-time windows, and $\delta$ is a small fixed amount of time, which is used as the length of each sliding-time window $\Delta=[t'-\delta, t']$. 

In essence, this metric assigns different weights to different time points of a given time interval, inversely proportional to their distance from t (i.e., smaller distance is associated with higher
weight). Specifically, it performs the following: it (1) breaks down the given time interval into shorter time periods using a sliding time window, (2) deploys an exponential decay function of the form $f(x)=e^{-\alpha x}$ to account for the time-decay effect, and (3) takes average of the causality values computed over each sliding time window~\cite{alvari2018early}.

\subsection{Semi-Supervised Causal Inference}
Having defined the causality-based features, we now proceed to present the proposed semi-supervised Laplacian SVM framework, \textsc{SemiPsm}. For the rest of the discussion, we shall assume a set of $l$ labeled pairs $\{(x_i,y_i)\}_{i=1}^l$ and an unlabeled set of $u$ instances $\{x_{l+i}\}_{i=1}^u$, where $x_i\in\mathbb{R}^n$ denotes the causality vector $\xi_{k}^I(i)$ of user $i$ and $y_i\in\{+1,-1\}$ (PSM or not).

Recall for the standard soft-margin support vector machines, the following optimization problem is solved:
\begin{equation}
\min_{f_\theta\in \mathcal{H}_k} \gamma||f_\theta||_k^2 + C_l \sum_{i=1}^{l}H_1(y_if_\theta(x_i))
\end{equation} 

In the above equation, $f_\theta(\cdot)$ is a decision function of the form $f_\theta(\cdot)=w.\mathbf{\Phi}(\cdot)+b$ where $\theta=(w,b)$ are the parameters of the model, and $\mathbf{\Phi(\cdot)}$ is the feature map which is usually implemented using the kernel trick~\cite{cortes1995support}. Also, the function $H_1(\cdot)=\max(0,1-\cdot)$ is the Hinge Loss function. The classical Representer theorem~\cite{belkin2005manifold} suggests that solution to the optimization problem exists in a Hilbert space $\mathcal{H}_k$ and is of the form $f_\theta^*(x) = \sum_{i=1}^{l}\alpha_i^*\mathbf{K}(x,x_i)$. Here, $\mathbf{K}$ is the $l\times l$ Gram matrix over labeled samples. Equivalently, the above problem can be written as:

\begin{align}
&\min_{w,b,\epsilon} \frac{1}{2}||w||_2^2 + C_l \sum_{i=1}^{l}\epsilon_i \\
&~s.t.~~~y_i(w.\mathbf{\Phi}(x_i)+b)\geq 1-\epsilon_i, ~ i=1,...,l \nonumber\\
&~~~~~~~~\epsilon_i\geq 0, ~ i=1,...,l
\end{align}

Next, we will use the above optimization equation as our basis to derive the formulations for our proposed semi-supervised learner.

The basic assumption behind semi-supervised learning methods is to leverage unlabeled instances in order to restructure hypotheses during the learning process~\cite{alvari2017semi}. Here, exogenous information extracted from causality-based features of users is exploited to make a better use of the unlabeled examples. To do so, we first introduce matrix $\mathbf{F}$ over both of the labeled and unlabeled samples with $\mathbf{F}_{ij}=||\mathbf{\Phi}(x_i)-\mathbf{\Phi}(x_j)||_2$ in $||.||_2$ norm. This way, we force instances $x_i$ and $x_j$ in our dataset to be relatively `close' to each other~\cite{beigi2018similar}, i.e., having a same label, if their corresponding causal-based feature vectors are close. To account for this, a regularization term is added to the standard equation and the following optimization is solved: 

\begin{equation}
\min_{f_\theta \in \mathcal{H}_k} \frac{1}{2}\sum_{i=1}^{l}\textbf{F}_{ij}||f_\theta(x_i)-f_\theta(x_j)||_2^2 = \mathbf{f}^T_\theta\mathcal{L}^T\mathbf{f}_\theta
\end{equation} 

\noindent where $\mathbf{f}=[f(x_1), ..., f(x_{l+u})]^T$ and $\mathcal{L}$ is the Laplacian matrix based on $\mathbf{F}$ given by $\mathcal{L}=\mathbf{D}-\mathbf{F}$, and $\mathbf{D}_{ii}=\sum_{j=1}^{l+u}\mathbf{F}_{ij}$. The intuition here is that causal pairs are more likely to have same labels than others.

%\begin{equation}
%\min_{f_\theta \in \mathcal{H}_k} %\frac{1}{2}\sum_{i=1}^{l}\textbf{A}_{ij}||f_\theta(x_i)-f_\theta(x_j)||_2^2 = %\mathbf{f}^T_\theta\mathcal{L}'^T\mathbf{f}_\theta
%\end{equation}

%We construct $\mathbf{G}$ with $(l+u)$ nodes in $\mathcal{F}_2$, and by adding an edge between each pair of nodes $\langle i,j \rangle$, if the edge weight $W_{ij}$ exceeds a given threshold. For computing the edge weights, we use the heat kernel~\cite{grigor2006heat} as a function of the Euclidean distance between two samples in $\mathcal{F}_2$, hence we set $W_{ij}=\exp^{-||x_i-x_j||^2/4t}$.

Following the notations used in~\cite{belkin2006manifold} and by including our regularization term, we would extend the standard equation by solving the following optimization:

\begin{equation}\label{eq:opt1}
\min_{f_\theta\in \mathcal{H}_k} \gamma||f_\theta||_k^2 
+
C_l \sum_{i=1}^{l}H_1(y_if_\theta(x_i)) 
+ 
C_r\textbf{f}_\theta^T\mathcal{L}\textbf{f}_\theta
\end{equation}

%Note one typical value for the smoothness penalty coefficient $C_s$ is $\frac{\gamma_I}{(l+u)^2}$, where $\frac{1}{(l+u)^2}$ is a natural scale factor for empirical estimate of the Laplace operator and $\gamma_I$ is a regularization term~\cite{belkin2006manifold}. 
Again, solution in $\mathcal{H}_k$ would be in the following form $f_\theta^*(x) = \sum_{i=1}^{l+u}\alpha_i^*\mathbf{K}(x,x_i)$. Here $\mathbf{K}$ is the $(l+u)\times(l+u)$ Gram matrix over all samples. The Eq.~\ref{eq:opt1} could be then written as follows:

\begin{align}
&\min_{\alpha,b,\epsilon} \frac{1}{2}\alpha^T\mathbf{K}\alpha + C_l \sum_{i=1}^{l}\epsilon_i + \frac{C_r}{2}\alpha^T\mathbf{K}\mathcal{L}\mathbf{K}\alpha
\end{align}

\begin{align}
&~s.t.~~~y_i(\sum_{j=1}^{l+u}\alpha_j\mathbf{K}(x_i,x_j)+b)\geq 1-\epsilon_i, ~ i=1,...,l 
\nonumber\\
&~~~~~~~~\epsilon_i\geq 0, ~ i=1,...,l
\end{align}

With introduction of the Lagrangian multipliers $\beta$ and $\gamma$, we write the Lagrangian function of the above equation as follows:

\begin{align}
L(\alpha,\epsilon,b,\beta,\gamma)=\frac{1}{2}\alpha^T\mathbf{K}(I+C_r\mathcal{L})\alpha
+C_l\sum_{i=1}^{l}\epsilon_i\\ \nonumber
-\sum_{i=1}^{l}\beta_i(y_i(\sum_{j=1}^{l+u}\alpha_j\mathbf{K}(x_i,x_j)+b)-1+\epsilon_i) - \sum_{i=1}^{l}\gamma_i\epsilon_i
\end{align} 

Obtaining the dual representation, requires taking the following steps:

\begin{align}
\frac{\partial L}{\partial b} = 0 \rightarrow \sum_{i=1}^{l}\beta_iy_i = 0 \\
\frac{\partial L}{\partial \epsilon_i} = 0 \rightarrow C_l - \beta_i - \gamma_i = 0 \rightarrow 0\leq\beta_i\leq C_l
\end{align}

With the above equations, we formulate the reduced Lagrangian as a function of only $\alpha$ and $\beta$ as follows:

%\begin{align}
%L^R(\alpha,\beta)=\frac{1}{2}\alpha^T\mathbf{K}(I+C_r\mathcal{L}+\frac{\gamma_I}{(l+u)^2}\mathcal{L}')\alpha
%\nonumber\\
%-\sum_{i=1}^{l}\beta_i(y_i(\sum_{j=1}^{l+u}\alpha_j\mathbf{K}(x_i,x_j)+b)-1+\epsilon_i)\nonumber\\
%\end{align} 

%This equation is further simplified as follows:

\begin{align}\label{eq:lagrange}
L^R(\alpha,\beta)= \frac{1}{2}\alpha^T\mathbf{K}(I+C_r\mathcal{L})\alpha
-\alpha^T\mathbf{K}\mathbf{J}^T\mathbf{Y}\beta+\sum_{i=1}^{l}\beta_i
\end{align} 

In the above equation, $\mathbf{J}=[\mathbf{I}~\mathbf{0}]$ is a $l\times(l+u)$ matrix, $\mathbf{I}$ is the $l\times l$ identity matrix and $\mathbf{Y}$ is a diagonal matrix consisting of the labels of the labeled examples. We first take the derivative of $L^R$ with respect to $\alpha$ and then set $\frac{\partial L^R(\alpha,\beta)}{\partial \alpha} = 0$. We have the following equation:

\begin{align}
\mathbf{K}(I+C_r\mathcal{L})\alpha
-\mathbf{K}\mathbf{J}^T\mathbf{Y}\beta = 0 
\end{align}

Accordingly, we obtain $\alpha^*$ by solving the following equation: 

\begin{align}\label{eq:alpha_star}
\alpha^* = (I+C_r\mathcal{L})^{-1}\mathbf{J}^T\mathbf{Y}\beta^*
\end{align}

Next, we obtain the dual problem in the form of a quadratic programming problem by substituting $\alpha$ back in the reduced Lagrangian function Eq.~\ref{eq:lagrange}:

\begin{align}\label{eq:beta_star}
\beta^* = {\operatorname{argmax}}_{\beta \in \mathbb{R}^l}~-\frac{1}{2}\beta^T\mathbf{Q}\beta + \sum_{i=1}^{l}\beta_i \\
s.t.~~~~
\sum_{i=1}^{l}\beta_iy_i = 0 \nonumber\\
0\leq \beta_i \leq C_l
\end{align}

\noindent where $\beta = [\beta_1,...,\beta_l]^T \in \mathbb{R}^l$ are the Lagrangian multipliers and $\mathbf{Q}$ is obtained as follows:

\begin{equation}\label{eq:Q}
\mathbf{Q} = \mathbf{YJK}(I+(C_r\mathcal{L})\mathbf{K})^{-1}\mathbf{J}^T\mathbf{Y}
\end{equation}

We summarize the proposed semi-supervised framework in Algorithm 1. Our optimization problem is very similar to the standard optimization problem solved for SVMs, hence we use a standard optimizer for SVMs to solve our problem. 

\begin{algorithm}
	\caption{\textbf{Semi-Supervised Causal Inference for PSM detection (\textsc{SemiPsm})}}\label{alg:alg1}
	
	\begin{algorithmic}[1]
		\Require $\{(x_i,y_i)\}_{i=1}^l$, $\{x_{l+i}\}_{i=1}^u$, $\mathcal{F}_1$, $\mathcal{F}_2$, $C_l$, $C_r$.
		\Ensure Estimated function $f_\theta: \mathbb{R}^n\rightarrow\mathbb{R}$
		\State \text{Construct matrix $\mathbf{F}$ based on the causality-based features}
		\State \text {Compute the corresponding Laplacian matrix $\mathcal{L}$.}
		\State \text{Construct the Gram matrix over all examples using $\mathbf{K}_{ij}=k(x_i,x_j)$} where $k$ is a kernel function.
		\State Compute $\alpha^*$ and $\beta^*$ using Eq.~\ref{eq:alpha_star} and Eq.~\ref{eq:beta_star} and a standard QP solvers.
		\State Compute function $f_\theta^*(x) = \sum_{i=1}^{l+u}\alpha_i^*\mathbf{K}(x,x_i)$
	\end{algorithmic}
\end{algorithm}

\subsection{Computational Complexity} Here, we will explain the scalability of the algorithm in terms of big-$\mathcal{O}$ notation for both constituents of the proposed framework separately. For the first part of the approach, given a set of $\mathcal{A}$ cascades, and average number of $avg(\tau)$ users' actions (i.e., timestamps) in each cascade where $\tau\in\mathcal{A}$, the complexity of computing causality scores is $\mathcal{O}(\mathcal{|A|}.(avg(\tau))^2)$ (note on average there are $(avg(\tau))^2$ pairs of users in each cascade). For the second part, i.e., learning the semi-supervised algorithm, the most time-consuming part is calculating the inverse of a dense Gram matrix which leads to ${\mathcal{O}((l+u)^3)}$ complexity, where $l$ and $u$ are number of labeled and unlabeled instances~\cite{belkin2006manifold}.

\section{Experiments}
In this section we conduct experiments on a Twitter ISIS-related dataset and present results for several supervised and semi-supervised approaches. We first explain the dataset and provide some data analysis. Then, we will present the baseline methods. Finally, results and discussion are provided.

\subsection{ISIS Twitter Dataset}
\begin{figure}[t]\center
	\includegraphics[width=0.34\textwidth]{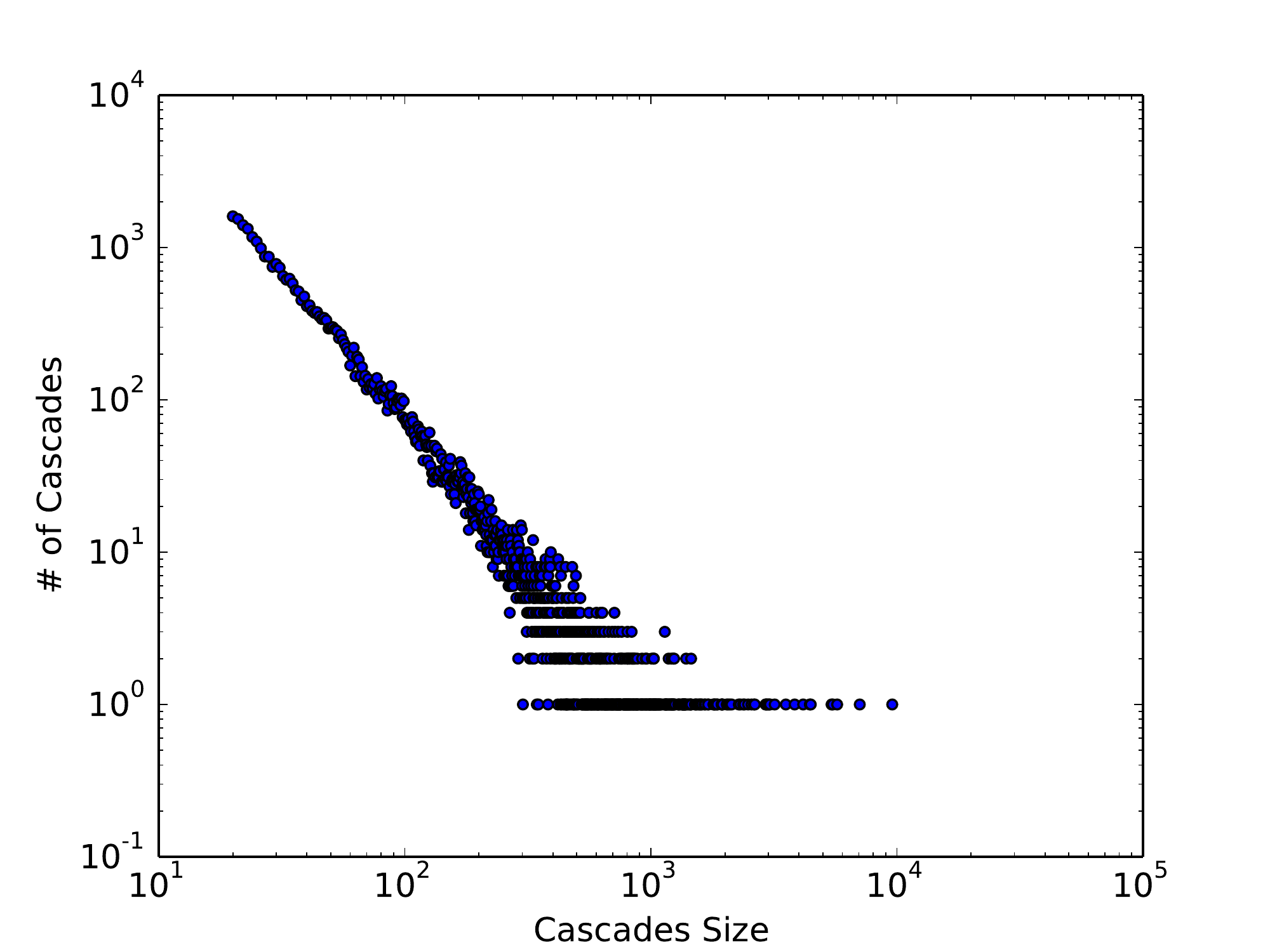}
	\includegraphics[width=0.34\textwidth]{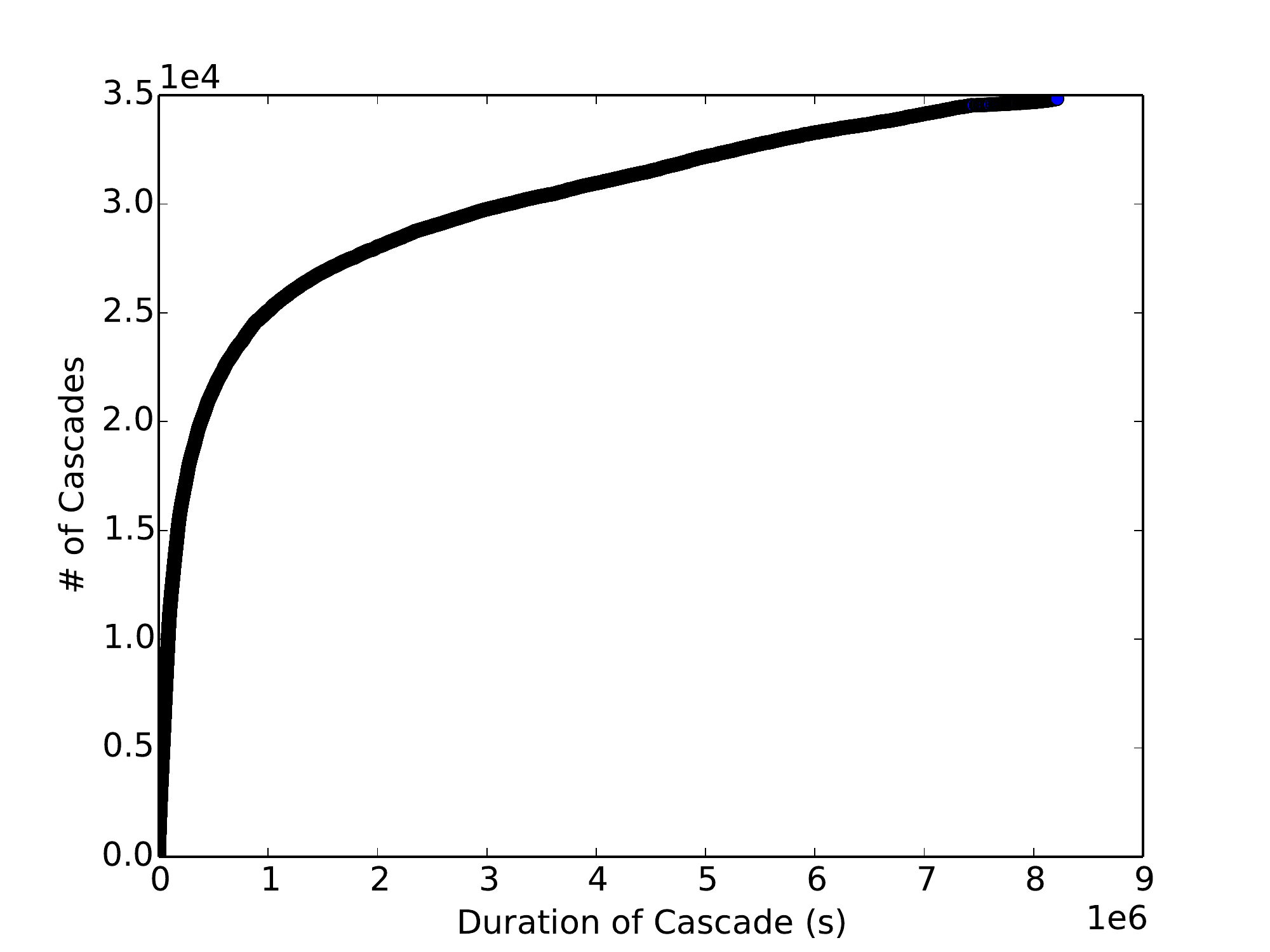}	
	\caption{(Top) Log-log distribution of cascades vs. cascade size. (Bottom) Cumulative distribution of duration of cascades.}
	\label{fig:data_analysis1}
\end{figure}

We collect a dataset (Table~\ref{tb:st}) of 53 M ISIS related tweets/retweets in Arabic, from Feb 22, 2016 to May 27, 2016. %We use the Twitter streaming API\footnote{https://developer.twitter.com/en/docs} which provides 1\% random tweets from the total volume of tweets at a particular moment. 
The dataset has different fields including user ID, retweet ID, hashtags, content, posting time. %The dataset also contains user profile information including name, number of followers/followees, description, location, etc. 
The tweets were collected using 290 different hashtags such as \textsf{\#Terrorism} and \textsf{\#StateOfTheIslamicCaliphate}. We use a subset of this dataset which contains 35 K cascades of different sizes and durations. There are $\sim$11 M tweets/retweets associated with the cascades. After pre-processing and removing duplicate users from cascades, cascades sizes (i.e. number of associated postings) vary between 20 to 9,571 and take from 10 seconds to 95 days to finish. The log-log distribution of cascades vs. cascade size and the cumulative distribution of duration of cascades are depicted in Figure~\ref{fig:data_analysis1}. 

\begin{figure}[t]\center	
	\includegraphics[width=0.34\textwidth]{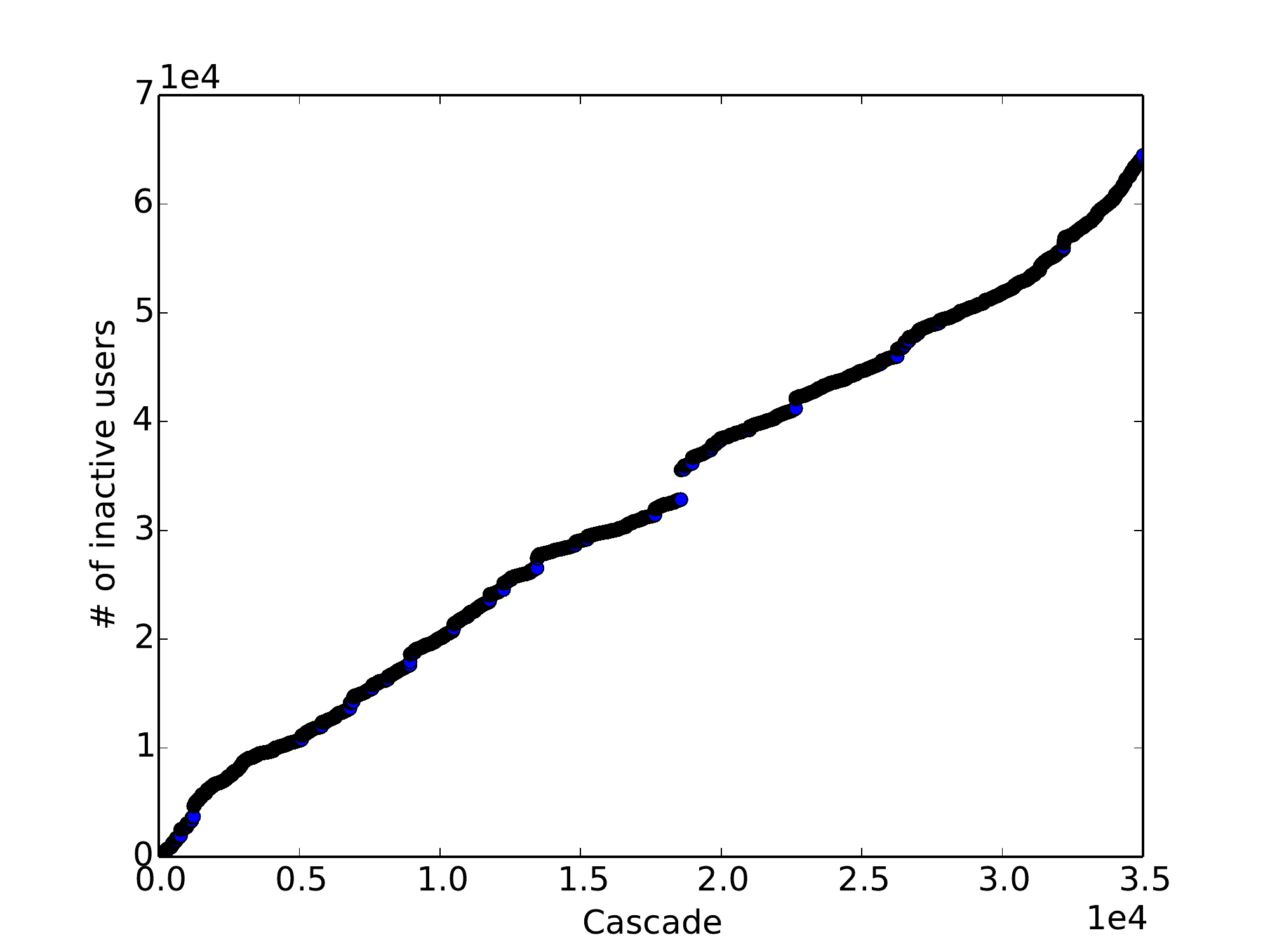}
	\caption{Total inactive users in each cascade.}
	\label{fig:data_analysis2}
\end{figure}

Based on the content of tweets in our dataset, PSMs are terrorism-supporting accounts who have participated in viral cascades. We chose to use threshold $\theta=100$ and take about 6 K viral cascades with at least 100 tweets/retweets. We demonstrate in Figure~\ref{fig:data_analysis2}, the total number of users in each cascade suspended by Twitter. %We experiment the effectiveness of our proposed approach on subsets of the training set with different sizes. Note we use no more than 50\% of original dataset to ensure our approach is able to identify PSMs early enough. 
We note that he dataset does not have any underlying network. We only focus on the non-textual information in the form of an \textit{action log}. We set $\phi=0.5$ to select \textit{key users}, i.e., we are looking for the users that participate in the cascades before the number of participants gets twice. After the data collection, we follow~\cite{thomas2011suspended} and check through Twitter API whether users have been suspended (PSM) or are still active (normal). According to Table~\ref{tb:st}, less than 24\% of the users in our dataset are PSM and the rests are normal.

\begin{table}[t]
	\centering
	\caption{Description of the dataset.}
	\begin{tabular}{|l|c|c|}
		\cline{1-3}
		\textbf{Name}          & \multicolumn{2}{c|}{\textbf{Value}}\\
		\hhline{===}
		\# of Cascades      & \multicolumn{2}{c|}{35 K}  \\ \cline{1-3}
		\# of Viral Cascades & \multicolumn{2}{c|}{6,602} \\ \cline{1-3}
		\# of Tweets/Retweets & \multicolumn{2}{c|}{10,823,168} \\ \cline{1-3}
		\# of Users  & PSM & Normal \\ \cline{2-3}
		& 19,859
		& 65,417\\
		\cline{1-3}
	\end{tabular}
	\label{tb:st}
\end{table}

\subsection{Baseline Methods}
We compare the proposed method \textsc{SemiPsm} against the following baseline methods. Note for all methods, we only report results when their best settings are used. 
\begin{itemize}
	\item \textbf{\textsc{LabelSpreading (Rbf Kernel)}~\cite{zhou2004learning}.} This is a graph inference-based label spreading approach with radial basis function (RBF) kernel.
	
	\item \textbf{\textsc{Label Spreading (Knn Kernel)}~\cite{zhou2004learning}.} Similar to the previous approach with K-nearest neighbor (KNN) kernel.
	
	\item \textbf{\textsc{Lstm}~\cite{kudugunta2018deep}.} The word-level LSTM approach here is similar to the deep neural network models used for sequential word predictions. We adapt the neural network to a sequence classification problem where the inputs are the vector of words in each tweet and the output is the predicted label of the tweet. We first use the word2vec~\cite{mikolov2013distributed} embedding pre-trained from a set of tweets similar to the data representation in our Twitter dataset. %The powerful concept behind word2vec is that word vectors that are close to each other in the vector space represent words that are not only of the same meaning but of the same context as well. 
		
	\item \textbf{\textsc{Account-Level (Rf Classifier)}~\cite{kudugunta2018deep}} This approach uses the following features of the user profiles: \textit{Statuses Count, Followers Count, Friends Count, Favorites Count, Listed Count, Default Profile, Geo Enables, Profile Uses Background Image, Verified, Protected}. We chose this method over Botometer~\cite{varol2017online} as it achieved comparable results with far less number of features (\cite{varol2017online} uses over 1,500 features)(see also~\cite{ferrara2016rise}). According to~\cite{kudugunta2018deep}, we report the best results when Random Forest (RF) is used.   	
	
	\item \textbf{\textsc{Tweet-Level (Rf Classifier)}~\cite{kudugunta2018deep}.} Similar to the previous baseline, this method uses only a handful of features extracted from tweets: \textit{retweet count, reply count, favorite count, number of hashtags, number of URLs, number of mentions}. Likewise, we use RF as the classification algorithm.

	\item \textbf{\textsc{SentiMetrix}~\cite{7490315}.} This approach was proposed by the top-ranked team in the DARPA Twitter Bot Challenge. We consider all features that we could extract from our dataset. Our features include tweet syntax (average number of hashtags, average number of user mentions, average number of links, average number of special characters), tweet semantics (LDA topics), and user behaviour (tweet spread, tweet frequency, tweet repeats). The proposed approach starts with a small seed set and propagates the labels. Since we have enough labeled data for the training part, we use Random Forest as the learning approach.
	
	\item \textbf{\textsc{C2dc}~\cite{alvari2018early}.} This approach uses time-decay causal community detection-based classification to detect PSM accounts~\cite{alvari2018early}. For community detection, this approach uses Louvain algorithm.
\end{itemize}

\subsection{Results and Discussion}
%\begin{table}[t]\small
%	\centering
%	\caption{Classification performance of various methods on the labeled data. For semi-supervised learners, the size of unlabeled data is fixed to 10\% of the training set. The best performance is in bold.}%\vspace{-2mm}
%	\begin{tabular}{|l|c|c|} 
%		\cline{1-3}
%		\textbf{Learner} & \textbf{F1-score} &
%		\textbf{AUC}
%		\\
%		\hhline{===}
%		\bf{\textsc{SemiPsm (Causal Features)}} & \bf{0.94} & \bf{0.88} \\ \hline
%		\bf{\textsc{SemiPsm (Account-Level Features)}} & 0.89 & 0.79\\ \hline
%		\bf{\textsc{SemiPsm (Tweet-Level Features)}} & 0.88 & 0.77\\ \hline
%		\bf{\textsc{LabelSpreading (Knn/Causal Features)}} & 0.89 & 0.69 \\ \hline
%		\bf{\textsc{LabelSpreading (Rbf/Causal Features)}} & 0.88 & 0.57 \\ \hline
%		\bf{\textsc{Lstm}} & 0.41 & 0.62 \\ \hline
%		\bf{\textsc{Account-Level (Rf Classifier)}} & 0.88 & 0.57 \\ \hline
%		\bf{\textsc{Tweet-Level (Rf Classifier)}} & 0.82 & 0.53 \\ \hline
%		\bf{\textsc{SentiMetrix}} & 0.54 & 0.74 \\ \hline
%		\bf{\textsc{C2dc}} & 0.39 & 0.6\\ \hline	
%	\end{tabular}
%	\label{tb:f1_auc_all}
%\end{table}

\begin{table}[t]
	\centering
	\caption{F1-score results of various methods on the labeled data. For semi-supervised learners, the size of the unlabeled data is fixed to 10\% of the training set. The best performance is in bold.}%\vspace{-2mm}
	\begin{tabular}{|l|c|} 
		\cline{1-2}
		\textbf{Learner} & \textbf{F1-score}
		\\
		\hhline{==}
		\bf{\textsc{SemiPsm (Causal Features)}} & \bf{0.94} \\ \hline
		\bf{\textsc{SemiPsm (Account-Level Features)}} & 0.89 \\ \hline
		\bf{\textsc{SemiPsm (Tweet-Level Features)}} & 0.88 \\ \hline
		\bf{\textsc{LabelSpreading (Knn/Causal Features)}} & 0.89 \\ \hline
		\bf{\textsc{LabelSpreading (Rbf/Causal Features)}} & 0.88 \\ \hline
		\bf{\textsc{Account-Level (Rf Classifier)}} & 0.88 \\ \hline
		\bf{\textsc{Tweet-Level (Rf Classifier)}} & 0.82 \\ \hline
		\bf{\textsc{SentiMetrix}} & 0.54 \\ \hline
		\bf{\textsc{Lstm}} & 0.41 \\ \hline	
		\bf{\textsc{C2dc}} & 0.4 \\ \hline			
	\end{tabular}
	\label{tb:f1_auc_all}
\end{table}

%\begin{table*} [t]  \centering
%	\caption{Classification performance of the semi-supervised approaches when causality-based features are used. Results are reported on different portions of the unlabeled data. The best performance is in bold.}
%	\begin{tabular}{@{} clcc|cc|cc|cc|cc @{}}
%		& &  \multicolumn{2}{c}{\textbf{10\%}}& \multicolumn{2}{c}{\textbf{20\%}} & \multicolumn{2}{c}{\textbf{30\%}} & \multicolumn{2}{c}{\textbf{40\%}} & \multicolumn{2}{c}{\textbf{50\%}}\\[2ex]
%		& & {\bf{{\footnotesize F1-score}}} & {\bf{{\footnotesize AUC}}}& {\bf{{\footnotesize F1-score}}} & {\bf{{\footnotesize AUC}}} & {\bf{{\footnotesize F1-score}}} & {\bf{{\footnotesize AUC}}} & {\bf{{\footnotesize F1-score}}} & {\bf{{\footnotesize AUC}}} &
%		{\bf{{\footnotesize F1-score}}} & {\bf{{\footnotesize AUC}}}\\
%		\cmidrule{2-12}
%		%& \bf{\textsc{SemiPsm}}  & \bf{0.94} & \bf{0.88} & \bf{0.93} & \bf{0.86} & \bf{0.91} & \bf{0.85} & \bf{0.9} & \bf{0.84} & \bf{0.88} & \bf{0.81}\\
%		& \bf{\textsc{LabelSpreading (Knn)}} & 0.89 & 0.69 & 0.88 & 0.66 & 0.87 & 0.65 & 0.85 & 0.62 & 0.81 & 0.58\\
%		& \bf{\textsc{LabelSpreading (Rbf)}} & 0.88 & 0.57 & 0.86 & 0.54 & 0.85 & 0.53 & 0.82 & 0.52 & 0.8 & 0.5\\
%		\cmidrule[1pt]{2-12}
%	\end{tabular}
%\label{tb:f1_auc_semi}
%\end{table*}

\begin{table} [t]  \centering
	\caption{F1-score results of the semi-supervised approaches when causality-based features are used. Results are reported on different portions of the unlabeled data. The best performance is in bold.}
	\begin{tabular}{@{} clc|c|c|c|l @{}}
		& &  \multicolumn{5}{c}{\textbf{Percentage of Unlabeled Data}}\\[2ex]
		& & {\bf{10\%}} & {\bf{20\%}} & {\bf{30\%}} & {\bf{40\%}} & {\bf{50\%}} \\
		\cmidrule{2-7}
		& \bf{\textsc{SemiPsm}}  & \bf{0.94} & \bf{0.93} & \bf{0.91} & \bf{0.9} & \bf{0.88}\\
		& \bf{\textsc{LabelSpreading (Knn)}} & 0.89 & 0.88 & 0.87 & 0.85 & 0.81\\
		& \bf{\textsc{LabelSpreading (Rbf)}} & 0.88 & 0.86 & 0.85 & 0.82 & 0.80 \\
		\cmidrule[1pt]{2-7}
	\end{tabular}
	\label{tb:f1_auc_semi}
\end{table}

All experiments were implemented in Python 2.7x and run on a machine equipped with an Intel(R) Xeon(R) CPU of 3.50 GHz with 200 GB of RAM running Linux. The proposed approach was implemented using CVXOPT\footnote{http://cvxopt.org/} package. Furthermore, we split the whole dataset into 50\% training and 50\% test sets for all experiments. We report results in terms of F1-score in tables~\ref{tb:f1_auc_all} and~\ref{tb:f1_auc_semi}. For any approach that requires special tuning of parameters, we conducted grid search to choose the best set of parameters. Specifically, for the proposed approach, we set the penalty parameter as $C_l=0.6$ and the regularization parameter $C_r=0.2$, and used linear kernel. For \textsc{LabelSpreading (Rbf)}, the default vale of $\gamma=20$ was used and for \textsc{LabelSpreading (Knn)}, number of neighbors was set to 5. Also, for random forest we used 200 estimators and the `entropy' criterion was used. For computing $k$ nearest neighbors in \textsc{C2dc}, we set $k=10$. 
%Finally, for LSTM, we use the skip-gram technique for the obtaining the word vectors where the input is the target word, while the outputs are the words surrounding the target words. We use a dimension 0f 50 for the hidden layer of the neural network trained on this input and output. For preprocessing the data, we remove stop words from the tweet sequence as well as hashtags which are not in the list of pre-trained embeddings. We limit the size of the word array obtained from the tweets to 10 and pad tweets which do not have 10 words. We feed the sequence of word embeddings corresponding to a tweet to the LSTM that outputs a single 32- dimension vector that is then fed forward through 2 ReLU activated layers of size 128 and 64 to give the output. 

Furthermore for LSTM, we preprocessed the individual tweets in line with the steps mentioned in \cite{soliman2017aravec}. Since the content of the tweets are in Arabic, we replaced special characters that were present in the text with their Arabic counterparts if they were present. %Along the same lines, we cleaned some of the Arabic characters that were mistyped with their correct ones. To transform the tweets into a form suitable for LSTMs, as an embedding we use a pre-trained set of Word Vectors (Word2Vec) \cite{mikolov2013distributed} meant for Twitter data. For obtaining the word vector for the Arabic tokens in the tweets in our dataset, we used pre-trained vectors trained on a similar twitter repository \footnote{https://github.com/bakrianoo/aravec}.  
We used word vectors of dimensions 100 and deployed the skip-gram technique for obtaining the word vectors where the input is the target word, while the outputs are the words surrounding the target words. To model the tweet content in a manner that uses it to predict whether an account is PSM or not, we used Long Short Term Memory (LSTM) models \cite{hochreiter1997long}. %, a superior variant of Recurrent Neural Networks. Such recurrent models have been found effective for NLP tasks, given their ability to learn relationships in sequential data. 
For the LSTM architecture, we used the first 20 words in the tokenized Arabic text of each tweet and use padding in situations where the number of tokens in a tweet are less than 20. We used 30 units in the LSTM architecture (many to one). The output of the LSTM layer was fed to a dense layer of 32 units with ReLU activations. We added dropout regularization  following this layer to avoid overfitting and the output was then fed to a dense layer which outputs the category of the tweets. 

We depict in Table~\ref{tb:f1_auc_all} classification performance of all approaches on the labeled data. For the proposed framework \textsc{SemiPsm}, we examine three sets of features (1) causality-based features, (2) account-level features~\cite{kudugunta2018deep}; and (3) tweet-level features~\cite{kudugunta2018deep}. For the graph inference-based semi-supervised algorithms, i.e., \textsc{LabelSpreading (Rbf)} and \textsc{LabelSpreading (Knn)}, we only report results where causality-based features are used as they achieved best performance with them. As it is observed from the table, the best results in terms of F1-score belong to \textsc{SemiPsm} where causality-based features are used. The runner-up is \textsc{SemiPsm} with account-level features and the next best approach is \textsc{SemiPsm} where tweet-level features are deployed. This clearly demonstrates the significance of using manifold regularization in the Laplacian semi-supervised framework over using other semi-supervised methods, \textsc{LabelSpreading (Rbf)} and \textsc{LabelSpreading (Knn)}. 

We further note that the supervised classifier Random Forest using both of the account-level and tweet-level features and the whole labeled dataset achieve worse or comparable results to the semi-supervised learners. The fact that obtaining several tweet and account-level features is not trivial and do not necessarily lead to the best classification performance, motivates us to use semi-supervised algorithms which use less number of labeled examples, and yet achieve competing performance. We also obtain an F1-score of 0.41 when LSTM is used-- the poor performance of the this neural network model can be attributed to the raw Arabic text content. It suggests that the Arabic tokens as a representation might not be very informative about the category of accounts it has been generated from and some kind of weighting might be necessary before the LSTM module is used. %Also, in future we intend to use a deep neural network with additional regularization compared to the shallow network used here.

Also, Table~\ref{tb:f1_auc_semi} shows the classification performance of the semi-supervised approaches with causality-based features. The results are achieved using different portions of the unlabeled data, i.e., $\{10\%,20\%,30\%,40\%,50\%\}$ of the training set. As it is seen in the table, \textsc{SemiPsm} achieves the best performance on different portions of the unlabeled data compared to the other semi-supervised learners, while performances of all approaches deteriorate with increasing the percentage of the unlabeled data. Furthermore, \textsc{SemiPsm} still outperforms all other supervised methods as well as \textsc{Lstm} and \textsc{C2dc} when up to 50\% of the data has been made unlabeled.

\textbf{Observations. } Overall, this paper makes the following observations:
\begin{itemize}
	\item Among the semi-supervised learners used in this study, \textsc{SemiPsm} achieves the best classification performance suggesting the significance of using unlabeled instances in the form of manifold regularization. Manifold regularization is shown effective in boosting the classification performance, with three different sets of features confirming this.
	\item Causality-based features achieve the best performance via both Laplacian and graph inference-based semi-supervised settings. This lies at the inherent property of the causality-based features-- they are designed to show whether or not user $i$ exerts a causal influence on $j$. This is effective in capturing PSMs as they are key users in making a message viral.
	\item Compared to the supervised methods \textsc{Account-Level (Rf)} and \textsc{Tweet-Level (Rf)}, semi-supervised learners achieve either comparable or best results, suggesting promising results with less number of labeled examples.
	\item Among the supervised methods \textsc{Account-Level (Rf)} and \textsc{Tweet-Level (Rf)}, the former achieves higher F1-score indicating that account-level features are more useful in boosting the performance, although they are harder to obtain~\cite{kudugunta2018deep}. 
	\item Semi-supervised learners achieve best or comparable results with supervised learners, even with up to 50\% of the data made unlabeled. This clearly shows the superiority of using unlabeled examples over labeled ones. 
\end{itemize}

\section{Related Work} The explosive growth of the Web has raised numerous
security and privacy issues. Mitigating these concerns has
been studied from several aspects~\cite{Beigi2018HT,alvari2016non,Cao:2014:ULG:2660267.2660269,beigi2018privacy,Cui:2013:COP:2487575.2487639,beigi2014leveraging,broniatowski2018weaponized,beigi2019protecting,alvari2019extremism}. Our work is related to a number of research directions. Below, we will summarize some of the state-of-the-art methods in each category while highlighting their differences with our work.

\noindent \textbf{Identifying PSM accounts.} Compared to~\cite{Causal2017} which uses causal inference to detect PSM accounts, our work utilizes time-decay causal inference (using sliding-time window) which allows for early detection of PSM. In contrast to~\cite{alvari2018early} where a causal community detection algorithm is proposed to leverage communities of PSM accounts in order to achieve higher performance, our work proposes a semi-supervised causal inference algorithm that achieves reasonable performance using less labeled data by utilizing unlabeled data.

\noindent \textbf{Social Spam/Bot Detection.} %The closest research direction to my research is social spam/bot detection. 
%Previous work on spam/bot detection either assumed network/textual information is available~\cite{Goyal:2010,7490315,wu2017adaptive} or did not differentiate between types of bots~\cite{6921650}. However, we did not assume any underlying network structure or cascade path information, and our approach is specific to PSM accounts. For example, 
Recently, DARPA organized a Twitter bot challenge to detect ``influence bots''~\cite{7490315}. Among the participants, the work of~\cite{Cao:2014:ULG:2660267.2660269}, used similarity to cluster accounts and uncover groups of malicious users. The work of~\cite{varol2017online} presented a supervised framework for bot detection which uses more than thousands features. In a different attempt, the work of~\cite{ICWSM1715678} studied the problem of spam detection in Wikipedia using different spammers behavioral features. There also exist some studies in the literature that have addressed (1) differences between humans and bots~\cite{chu2012detecting}, (2) different natures of bots~\cite{varol2017online} or (3) differences between bots and human trolls~\cite{broniatowski2018weaponized}. For example the work of~\cite{chu2012detecting} conducted a series of measurements in order to distinguish humans from bots and cyborgs, in term of tweeting behavior, content, and account properties. To do so, they used more than 40 million tweets posted by over 500 K users. Then, they performed analysis and find groups of features that are useful for classifying users into human, bots and cyborgs. They concluded that entropy and certain account properties can be very helpful in differentiating between those accounts. In a different attempt, some other studies have tried to differentiate between several natures of bots. For instance, in the work of~\cite{varol2017online}, authors performed clustring analysis and revealed specific behavioral groups of accounts. Specifically, they identified different types of bots such as \textit{spammers}, \textit{self promoters}, and \textit{accounts that post content from connected applications}, using manual investigation of samples extracted from clusters. Their cluster analysis emphasized that Twitter hosts a variety of users with diverse behaviors; that is in some cases the boundary between human and bot users is not sharp, i.e. some account exhibit characteristics of both. 

Also, the work of~\cite{broniatowski2018weaponized}, uses Twitter data to quantify the impact of Russian trolls and bots on amplifying polarizing and anti-vaccine tweets. They first used the Botometer API to assign bot probabilities to the users in the dataset and divided the whole dataset into 3 categories: those with scores less than 20\% (very likely to be human), between 20\% and 80\% (e.g., cyborgs with uncertain provenance) and above 80\% (high likely to be bots). Then, they posed two research questions: (1) Are bots and trolls more likely to tweet about vaccines?, and (2) Are bots and trolls more likely to tweet polarizing and anti-vaccine content? Their analysis demonstrated that Twitter bots and trolls significantly impact on online discussion about vaccination and this differs by account type. For example, Russian trolls and bots post content about vaccination at higher rates compared to an average user. Also, according to this study, troll accounts and content polluters (e.g., dissemination of malware, unsolicited commercial content, etc.) post anti-vaccine tweets 75\% more than average users. In contrast, spambots which can be easily distinguished from humans, are less likely to promote anti-vaccine messages. Their closing remarks suggest strongly that distinguishing between malicious actors (bots, trolls, cyborgs, and human users) is difficult and thus anti-vaccine messages may be disseminated at higher rates by a combination of these malicious actors. 

In contrast to the above works, our work does not deploy any extra information (e.g., user-related attributes or network-based features) other than users' actions (i.e., cascade with timestamps). It is also worthwhile to note that most of the existing well-known bot detection algorithms such as Botometer~\cite{davis2016botornot} leverage over one thousand features in order to detect high-likely bots. %As a final note, these approaches can complement each other as I can easily incorporate these features into my approach. There is a comprehensive survey on the ongoing efforts to fight social bots in~\cite{ferrara2016rise}.

\noindent \textbf{Fake News Identification.} A growing body of research is addressing the impact of bots in manipulating political discussion, including the 2016 U.S. presidential election~\cite{shao2017spread} and the 2017 French election~\cite{ferrara2017}. For example,~\cite{shao2017spread} analyzes tweets following recent U.S. presidential election and found evidences that bots played key roles in spreading fake news.  

\noindent \textbf{Identifying Instigators.} There are some work on instigator detection~\cite{journals/corr/PeiMAZM14,Fu2015} and outbreak prediction~\cite{Cui:2013:COP:2487575.2487639}. In~\cite{Konishi:2016:IKO:3061053.3061145}, authors performed classification to detect users who adopt popular items. In~\cite{Zhu:2016:ISD:2942477.2942508}, authors designed an approach for information source detection and in particular initiator of a cascade. Our work is focused on a set of users who \textit{might} or \textit{might not} be initiators. Our work is different from these works since we leverage causality analysis to detect causes of popularity of messages that go viral. 

\noindent \textbf{Extremism and Water Armies Detection.} The work of~\cite{DBLP:journals/corr/KlausenMZ16} designed a behavioral model to detect extremists. Authors in~\cite{benigni2017online} performed iterative vertex clustering and classification to identify Islamic Jihadists on Twitter. The works of~\cite{DBLP:journals/corr/abs-1111-4297,wang2014detection} also used user behavioral and domain-specific attributes to detect water armies. Our work also differs from these works as we do not use any features such as network/user attributes. 

\noindent \textbf{Causal Reasoning.} As opposed to~\cite{DBLP:journals/corr/abs-1205-2634,DBLP:journals/corr/StantonTJVCS15,Kleinberg:2011:LCI:2283516.2283556} which deal with preconditions as single atomic propositions, we use rules with preconditions of more than one atomic propositions. %Also, neither of~\cite{DBLP:journals/corr/abs-1205-2634,Causal2017} have addressed early detection of PSMs.
\section{Conclusion}
We presented a semi-supervised Laplacian SVM to detect PSM users in social media who are promoters of misinformation spread. We cast the problem of identifying PSMs as an optimization problem and introduced a Laplacian semi-supervised SVM via utilizing unlabeled examples through manifold regularization. In this work, we examined different sets of features extracted from users activity log (in the form of cascades of retweets) as regularization terms: (1) causality-based features; and (2) LSTM-based features. Our causality-based features were built upon \textit{Suppes' theory of probabilistic causation}. The LSTM-based features were extracted via LSTM which has shown promising results for different tasks in the literature. 

In future, we would like to replicate the study by feeding other sets of features such as time-series features and those extracted using LSTM to the semi-supervised framework. Also, we plan to investigate other forms of causality inferences and other regularization terms to seek if we can further improve the classification performance by distinguishing between different types of PSMs.

\section{Acknowledgment}
Some of the authors are supported through the DoS and DoD Minerva program.

\bibliographystyle{ACM-Reference-Format}
%\bibliography{sample-bibliography}

\end{document}